# Information-Driven Fault Detection and Identification For Multi-Agent Spacecraft Systems: Collaborative On-Orbit Inspection Mission

Akshita Gupta[*]
*Purdue University, West Lafayette, IN, 47907, USA*

Arna Bhardwaj[†]
*Georgia Institute of Technology, Atlanta, GA, 30332, USA*

Yashwanth Kumar Nakka[‡]
*Georgia Institute of Technology, Atlanta, GA, 30332, USA*

Changrak Choi[§] and Amir Rahmani[¶]
*Jet Propulsion Laboratory, California Institute of Technology, Pasadena, CA, 91104, USA*

In this chapter, we present a global-to-local, task-aware fault detection and identification (FDI) framework for multi-spacecraft systems performing a collaborative inspection mission in low Earth orbit (LEO). The scenario considers multiple observer spacecraft in stable passive relative orbits (PROs) conducting visual inspection or mapping of a target spacecraft. The inspection task is encoded as a global cost functional, $\mathcal{H}$, which directly incorporates the inspection sensor model, each agent's full pose, and the mission's information-gain objective. This cost functional drives both global-level decision making (task allocation and orbit reconfiguration) and local-level actions (agent motion and sensing), enabling tight coupling between mission objectives and FDI. Fault detection is performed by comparing the expected and observed global task metric $\mathcal{H}$, using thresholds derived from the mission context. To enable fault identification, we introduce higher-order cost gradient metrics that discriminate between task-specific sensor faults, agent-level actuator faults, and state sensor faults. Furthermore, we propose an adaptive thresholding mechanism that accounts for the time-varying nature of the inspection task and the dynamic evolution of PRO-based observation geometries.

The proposed framework is validated in simulation for a representative multi-spacecraft collaborative inspection mission, demonstrating the reliable detection and classification of diverse fault types, including inspection sensor degradation and actuator malfunctions, while maintaining mission objectives. This approach unifies information-driven guidance and control with task-aware FDI, providing a pathway toward fault-resilient autonomous inspection architectures for future distributed spacecraft missions.

## Nomenclature

| | | |
|---|---|---|
| $\sigma(\mathbf{p}, \mathbf{s})$ | = | variance of estimating point of interest, $\mathbf{s}$ with a sensor at pose, $\mathbf{p}$ |
| $\boldsymbol{\tau}$ | = | torque (N m) |
| $\phi(\mathbf{s}) \in [0, 1]$ | = | relative importance of point of interest, $\mathbf{s}$ |
| $\psi(s)$ | = | consensus term |
| $\omega_{low}, \omega_{mid}, \omega_{high}$ | = | loop frequency of low-, mid-, and high-rate components (Hz) |
| $\boldsymbol{\omega}$ | = | angular velocity (rad s$^{-1}$) |

[*]Graduate Researcher, School of Industrial Engineering, gupta417@purdue.edu
[†]Graduate Researcher, School of Aerospace Engineering, abhardwaj82@gatech.edu
[‡]Assistant Professor, Director of Aerospace Robotics Lab, ynakka3@gatech.edu
[§]Robotics Technologist, Maritime and Multi-Agent Autonomy Group, Jet Propulsion Laboratory, Changrak.Choi@jpl.nasa.gov
[¶]Supervisor, Maritime and Multi-Agent Autonomy Group, Jet Propulsion Laboratory, Amir.Rahmani@jpl.nasa.gov

| | | |
|---|---|---|
| $\mathbf{A}$ | = | state transition matrix for continuous time |
| $\mathbf{A_d}$ | = | state transition matrix for discrete time |
| $\mathbf{A_{1q}, A_{1w}, A_{2w}}$ | = | state transition matrices of linearized attitude dynamics |
| $\mathbf{A_{1q}, A_{1w}, A_{2w}}$ | = | state transition matrices of linearized attitude dynamics |
| $\mathbf{B}$ | = | input matrix for continuous time |
| $\mathbf{B_d}$ | = | input matrix for discrete time |
| $d_{min}$ | = | minimum safe distance from the observer to the target spacecraft (m) |
| $d_s$ | = | desired sensing distance from the observer to the target spacecraft (m) |
| ECI | = | Earth Centered Inertial Frame |
| FoV | = | sensor field of view (rad) |
| $H$ | = | information cost |
| $\mathbf{J}$ | = | inertia matrix of the observer spacecraft ($\mathrm{kgm}^2$) |
| LVLH | = | Local-Vertical Local-Horizontal Frame |
| $m$ | = | mass of the observer spacecraft (kg) |
| $n$ | = | mean motion of the target spacecraft ($\mathrm{rads}^{-1}$) |
| $N$ | = | number of deputies (or) observers |
| $\mathbf{p}$ | = | pose of sensor on the observer spacecraft |
| $\mathcal{P}$ | = | set of all sensor poses |
| PRO | = | Passive Relative Orbit |
| POI | = | Points Of Interest on surface of the target spacecraft |
| $\mathbf{q}$ | = | attitude of the observer spacecraft as quaternion |
| $\bar{\mathbf{q}}$ | = | nominal attitude trajectory of the observer spacecraft |
| $\mathbf{s}$ | = | sampled point of interest on the target spacecraft |
| $\mathcal{S}_i(t)$ | = | set of visible POIs for spacecraft $i$ at time $t$ |
| $\tau_i(t)$ | = | adaptive fault threshold for spacecraft $i$ at time $t$ |
| $T$ | = | time step (s) |
| $\mathbf{u}$ | = | control input for relative orbit dynamics (N (or) N m) |
| $\mathcal{U}$ | = | convex control constraint set |
| $w$ | = | variance from prior model |
| $\mathbf{x}_i$ | = | state of the $i^{\text{th}}$ observer spacecraft |
| $\bar{\mathbf{x}}_i$ | = | nominal trajectory of the $i^{\text{th}}$ observer spacecraft |
| $(x, y, z)$ | = | relative orbit coordinates in LVLH frame |

# I. Introduction

Multi-spacecraft systems offer a new class of missions that are more flexible and adaptive than traditional single-spacecraft missions for near-Earth as well as deep-space applications. Several past missions have successfully utilized and demonstrated the benefits of such multi-spacecraft system architecture. For example, the Afternoon Constellation (A-Train) - a formation that included satellites such as Aqua [1], Aura [2], PARASOL [3], CloudSat [4], CALIPSO [5], GCOM-W1 [6], and OCO-2 [7], have leveraged the distributed simultaneous observations to study the Earth climate. By flying a train of heterogeneous instruments in close succession, the constellation enabled the ground-based fusion of near-simultaneous observations to create comprehensive, multi-dimensional views of Earth's climate system. On the other hand, missions such as GRACE [8] and GRACE-FO [9, 10] have launched two identical spacecraft flying in tandem as a single, distributed instrument. By precisely measuring minute changes in their relative separation, these missions created unprecedented maps of Earth's time-varying gravity field.

While the aforementioned missions demonstrated the immense scientific potential of distributed architectures, the interaction between the spacecraft in the system was minimal. For example, the multiple satellites in the Afternoon Constellation operated independently, and coordination among satellites was primarily a ground-managed station-keeping exercise to maintain a loose formation in the orbit. GRACE and GRACE-FO had a more tightly coupled system that performed tandem flying based on continuous, high-precision inter-satellite ranging. Nevertheless, both were following predefined tasks with no component of onboard collaborative autonomous decision-making that would act based on the state of other spacecraft.

In recent years, rapid progress in hardware miniaturization and intelligent autonomy [13] has sparked increased interest in developing multi-spacecraft missions with far more complex interactions and coordination among the spacecraft members. The upcoming missions, such as CADRE [14–16] and SunRISE [17, 18], significantly leverage the multi-spacecraft system architecture and high-level autonomy to enable science objectives through distributed observations. The Cooperative Autonomous Distributed Robotic Exploration (CADRE) mission, for example, will deploy a team of three suitcase-sized rovers to the lunar surface to demonstrate multi-agent autonomy. Using a hierarchical, on-board autonomy system, the rovers will cooperatively plan and execute tasks without direct human control and with minimal ground-in-the-loop interventions. A key demonstration will be a multi-static ground-penetrating radar (GPR) survey, where the rovers must drive in a precise geometric formation to create a 3D map of the lunar subsurface. In this modality, the rover team itself functions as a single, distributed scientific instrument, where the GNC system's ability to maintain the formation is directly coupled to the quality of the scientific data, exemplifying a tightly-coupled, task-aware system.

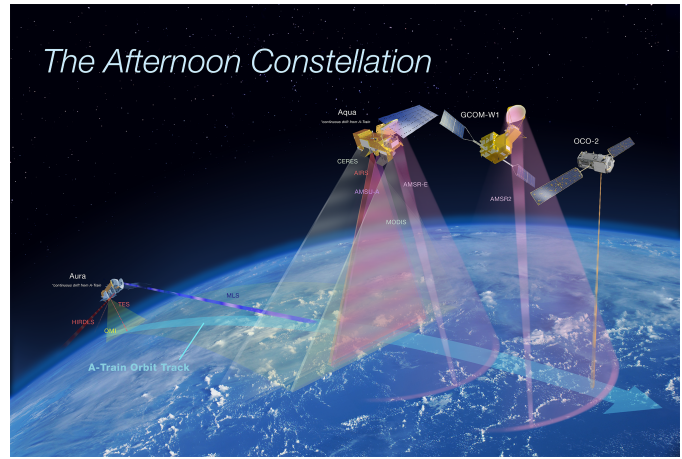

**Fig. 1** The Afternoon Constellation consisted of multiple satellites, including Aqua, Aura, CALIPSO, CloudSat, and OCO-2, that closely followed one another in the same orbital track (A-Train). The overlapping instrument swaths enabled synergistic, multi-dimensional observations of Earth's atmosphere, but the inter-spacecraft coordination was limited to ground-managed station-keeping within predefined orbital boxes [11] Credit: NASA.

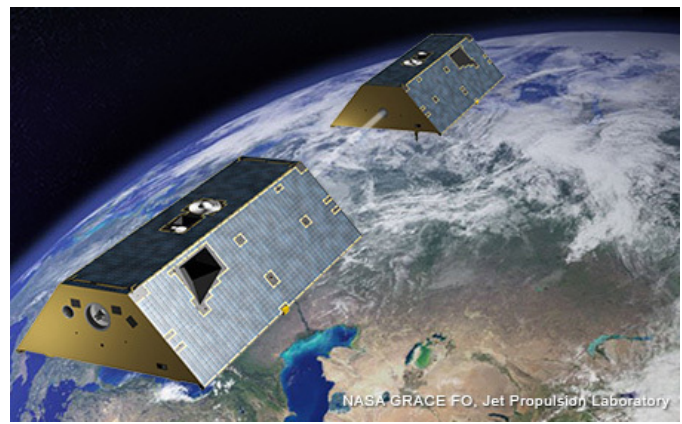

**Fig. 2** The GRACE-FO twin-satellite system. The mission measured changes in Earth's gravity field by precisely tracking minute variations in the distance between the two spacecraft using a microwave and laser ranging system. This architecture transformed the spacecraft pair into a single, distributed scientific instrument [12] Credit: NASA/JPL-Caltech.

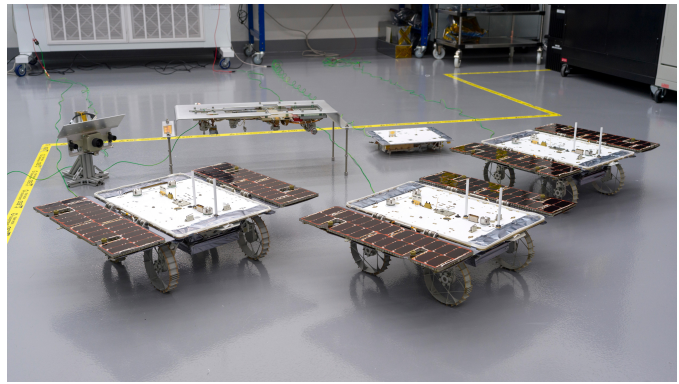

**Fig. 3   Three CADRE (Cooperative Autonomous Distributed Robotic Exploration) rovers in a clean room at NASA Jet Propulsion Lab. The mission will demonstrate multi-agent autonomy by having the rovers cooperatively perform tasks, such as driving in formation to create a 3D subsurface map with ground-penetrating radar. Credit: NASA/JPL-Caltech [19].**

In contrast, the Sun Radio Interferometer Space Experiment (SunRISE) mission will utilize a constellation of six 6U form factor CubeSats in a passive formation in Earth orbit to study solar activity. Together, these spacecraft will form a single, 10-km-wide virtual radio telescope through the technique of aperture synthesis. Each CubeSat acts as an independent antenna, and their data is combined on the ground to image low-frequency solar radio bursts—phenomena that are unobservable from Earth due to ionospheric blockage. This distributed architecture is the enabling technology for the mission, as it overcomes the fundamental physical limitation of building and deploying a monolithic 10-km structure in space. SunRISE exemplifies a loosely coupled system where complexity is shifted from onboard control to ground-based data processing and precise position knowledge.

While a multi-spacecraft mission architecture inherently offers robustness to faults through redundancy and the ability to reconfigure task allocation, this robustness is not absolute. The increased system-level complexity, arising from the tight coupling between multiple agents, their subsystems, and the shared mission objectives, creates numerous potential points of failure. Moreover, these systems depend on distributed communication and control architectures that, while enabling coordinated operation, can also act as conduits for fault propagation across the network. For example, in a leader–follower formation, a significant disturbance in the motion trajectory of even a single follower—caused by actuator degradation, sensor miscalibration, or environmental perturbations—can propagate through the control coupling between agents, leading to distortion of the formation geometry and loss of coordinated coverage. Similarly, in distributed sensing missions, a single agent exhibiting adversarial, faulty, or biased sensing behavior can corrupt the shared state estimates in neighboring agents through the underlying consensus or data-fusion framework, leading to systemic degradation in situational awareness. In extreme cases, such faults can create positive feedback loops in the estimation or control process, amplifying their effects and potentially leading to mission failure.

Additionally, the heterogeneous nature of multi-spacecraft systems—where agents may differ in sensing modalities, actuation capabilities, and onboard autonomy—means that faults can manifest in mission-specific ways, such as loss of a critical sensing mode, misalignment of a high-gain antenna, or degradation of relative navigation accuracy. Because mission performance often depends on the coordinated execution of tasks (e.g., maintaining stable passive relative orbits, synchronized imaging, or cooperative manipulation), even faults in a subset of agents can have disproportionate impacts on global mission objectives.

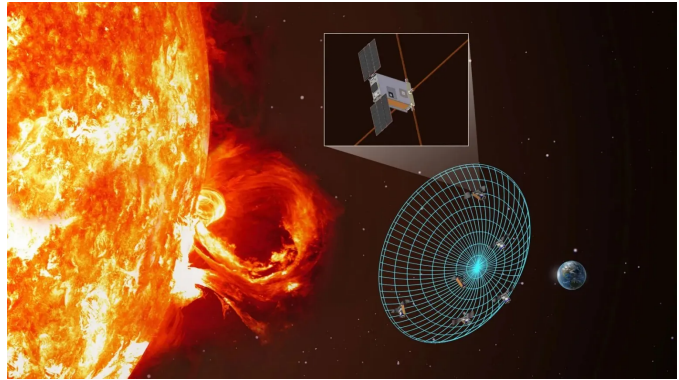

**Fig. 4 Artist's concept of the SunRISE (Sun Radio Interferometer Space Experiment) mission. Six 6U form factor CubeSats will fly in a passive formation, acting as a single, 10-km-wide virtual radio telescope to study low-frequency emissions from solar storms, which are unobservable from Earth [20] Credit: NASA.**

These vulnerabilities highlight the necessity for a Fault Detection, Isolation, and Recovery (FDIR) architecture that spans both the network and individual-agent levels. Such an architecture must rapidly detect anomalies, accurately identify their source and type (e.g., actuator, sensor, or task-specific performance faults), and execute recovery strategies—including re-tasking, dynamic reconfiguration of formations, or degradation of performance requirements—to ensure graceful mission degradation rather than abrupt failure. In this context, task-aware FDIR approaches, which link fault metrics directly to mission objectives, offer a principled way to maintain operational effectiveness even in the presence of persistent or cascading faults. To this end, we present an information-driven approach that incorporates both network-level task and agent-level system performance. In the following sections, we present our recent work, outlining its key contributions and situating its importance within the broader context of multi-spacecraft fault detection and autonomous mission resilience.

### A. Related Work

In the field of Fault Detection and Identification (FDI), a fault is defined as a deviation of the system from its nominal behavior [21]. Fault detection refers to identifying the occurrence of a fault, while fault identification determines the type and magnitude of the fault that has occurred. Faults cause anomalous behavior of the system and can lead to failure of the mission as the system's ability to perform a required function is interrupted. The primary goals of FDI algorithms are to detect faults in finite time and to precisely locate and identify them in order to prevent further propagation and system failure. At the level of individual agents, sensor, actuator, and system parameter degradation are commonly encountered types of faults [22]. Sensor faults involve the corruption of data from sensors such as star trackers, gyros, or GPS receivers. The corruption can manifest in various forms, including bias, drift, loss of precision, or even complete malfunction. Actuator faults lead to the inability of a spacecraft to correctly apply control actions, such as thruster lock-in-place or motor failure. System parameter degradation refers to changes in the physical properties of the system (e.g., resistance, stiffness, material stress) that lead to an altered response of the system to nominal control inputs. For multi-agent systems, propagation of fault data, packet delay, packet loss, and network link failure are common types of faults that can occur at the network level leading to fault propagation and cascade failure (even collision between the agents in the network).

FDI algorithms are typically developed by first modeling the nominal behavior of the system, then formulating a fault-sensitive metric and defining an associated threshold to distinguish between nominal and off-nominal conditions. In distributed multi-agent systems, the evolution of a given state variable depends not only on the system dynamics and control inputs but also on information received from neighboring agents. Within a consensus framework, all agents are required to agree on a set of predefined consensus variables to accomplish a shared global task. However, these consensus variables may not always be directly observable. For example, in online rendezvous and trajectory planning applications, agents may need to achieve consensus on mission-specific parameters, such as flight time [23, 24] in real-time, that are inherently partially observable and cannot be measured by external sensors. This challenge complicates the design of robust FDI algorithms, as faults must be inferred from indirect indicators rather than directly measured variables.

For observable state (or consensus) variables, most works in the literature develop FDI methods using state estimator–based approaches. For example, in [25], the authors proposed a local FDI algorithm for each agent to detect sensor faults. Their method employs a partition-based Luenberger estimator, where each agent estimates a specific component of the global state vector. Since the local dynamics are coupled with the states of neighboring agents, each agent computes a residual vector to determine whether a sensor fault has occurred. The detection threshold in this approach is defined as a function of the time-varying error covariance matrix. A more practical FDI algorithm was presented in [26], which accounts for measurement noise and communication delays. However, the use of a time-varying threshold in this method restricts detection to faults that lie outside the feasible set of state variables. In another approach, [27] applied Bayesian analysis to identify and exclude outlier measurements from the sensor network before performing FDI. In that work, the detection thresholds were determined empirically, rather than being derived from system dynamics or estimation uncertainty.

When state estimation is not a feasible approach for generating local residual vectors, FDI algorithms can instead be designed to analyze statistical trends in the data exchanged between agents. Fault and adversary detection methods in this category are generally applicable to a wide range of distributed optimization frameworks. Broadly, these methods can be classified into two categories: (i) approaches that exploit redundancy in network topology to guarantee resilience, and (ii) approaches that rely on statistical analysis of shared data to detect anomalies. In the first category, [28, 29] demonstrated that, in the presence of $k$ adversarial nodes, consensus can still be achieved if each regular node has at least $(2k + 1)$ neighbors. Their proposed algorithm implements a local filtering technique that discards information from $2k$ neighbors whose values lie at the extremes, thereby reducing the influence of adversarial agents. However, this method guarantees convergence only within the convex hull of the minimizers of the regular nodes. The local filtering concept has also been applied in distributed state estimation problems [30, 31]. Despite its theoretical guarantees, ensuring topological redundancy in practice is challenging due to increased communication overhead, especially in bandwidth-constrained environments. Moreover, the resilience of these methods does not scale proportionally with network size, as the allowable number of adversarial nodes does not grow with the total number of nodes [32, 33].

Given the limitations of topology-based methods, *analyzing the statistical trends of shared data* to identify adversarial nodes offers a more promising and scalable alternative. In this context, [34, 35] propose a *gradient-based metric* for detecting malicious agents. Specifically, [35] considers an attack scenario in which the local objective functions of adversarial nodes are arbitrarily modified, while [34] addresses *false data injection attacks*. In both cases, each node maintains a score for its neighbors by estimating their gradients over time and progressively severs links with neighbors whose scores consistently exceed those of the rest. The detection and isolation algorithms in [34, 35] are demonstrated empirically; however, despite providing intuitive justification for using a gradient-based metric, these works offer *no formal theoretical guarantees* on convergence. Furthermore, the resilient algorithm in [35] is designed for *directed graphs* and requires each node to share its data with exactly one neighbor per time step—a constraint that significantly slows convergence.

Statistical trend analysis has also been widely applied in the *distributed estimation* problem. For example, [36] develops a method for detecting faulty data injection in a *single-sensor system* using a *Chi-squared test* on the measurement innovation. Here, the attacker employs a *linear deception strategy* to preserve the statistical properties of the measurements, and the authors derive an optimal detection strategy for this setting. This work was later extended to the *multi-sensor* case [37, 38], considering a centralized remote state estimation system and proposing detection schemes for different attack scenarios. However, a key limitation of these methods is that they do not address how to *select or adapt the detection threshold* required to classify a node as adversarial [36–39].

To this end, a fault detection, isolation, and recovery (FDIR) architecture is essential for accommodating potential faults at both the network and individual agent levels, allowing the mission to continue with graceful degradation. In this work, we present a new FDI method, as shown in Fig. 5, that detects failures at the network level using an abstraction of the global task objective $\mathcal{H}$ and local sensing information, and informs the agent-level FDIR algorithm to perform the necessary actions for recovery. For example, a fault at the network level could be caused by communication loss, a global task sensor fault (for inspection tasks), a global task actuator fault (for on-orbit construction), or a fault at the agent level could be due to thruster or reaction wheel issues. We propose a simulation vs. real comparison using the $\mathcal{H}$ and its higher-order gradients. We detect the fault by computing the off-nominal behavior from the expected global task objective $\mathcal{H}$ by monitoring the individual task using a residual vector sensitive to the agent's faults. The global task objective $\mathcal{H}$ is designed to be a function of the state of agents in the network and a model of the task sensor (for inspection) or actuator (for construction). The residual vector is a function of the local relative state estimates of the agents in the network. We propose a metric that computes the deviation of the $\mathcal{H}$ from the nominal performance, which was defined from empirical simulations done on the system. In field operations the nominal performance is judged by the user, which is used as an indicator for faults at the network level, and uses higher-order derivatives of $\mathcal{H}$ to infer if the agent-level faults, as shown in Fig. 5.

**Global Task-Awareness**          **Fault Detection and Identification**

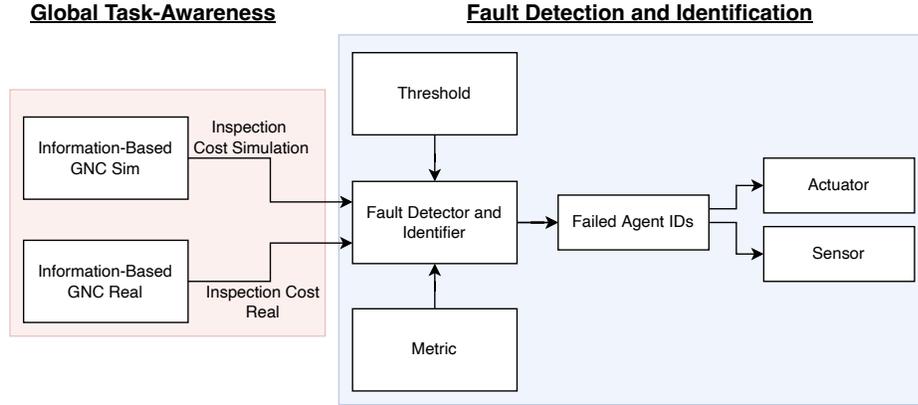

**Fig. 5    Global task-aware fault detection and isolation for a distributed spacecraft network.**

Prior work [27, 40] includes FDI architectures for distributed sensor networks that utilize a local decentralized observer to detect internal agent sensor or actuator failures. Recent work [41, 42] uses adaptive or reconfiguration control with a minimal notion of network task to achieve fault tolerance. We instead focus on incorporating global task objectives to inform the local FDIR, enabling it to react or respond appropriately and continue the mission autonomously to the best possible extent of the network's capability, while maximizing the global task objective $\mathcal{H}$. Furthermore, while most of the earlier work [25, 26] on FDI for distributed systems focuses on simple linear dynamical systems, we focus on low-earth orbit formation flying dynamics that includes periodic orbits.

**Main Contributions.** The main contributions of this work are as follows: 1) we propose an architecture for FDI in a multi-agent spacecraft system that integrates global-task objectives and local-agent level behaviors for task awareness, 2) we derive a global cost functional that is decomposable to cost functions that inform local progress and intermediate consensus on global progress, 3) we propose a novel FDI metric based on the global cost $\mathcal{H}$ and the high-order derivatives of the $\mathcal{H}$ to detect and identify both the global and local faults.

We apply our FDI architecture to a recently proposed multi-agent collaborative spacecraft inspection mission [43] in a low Earth orbit to detect failures in the inspection sensors and individual agent sensing. The cost function $\mathcal{H}$ defines the global inspection progress by fusing individual agent sensor data measurements. The inspection data fusion runs at a fixed frequency $\omega_g$, and the network fault diagnosis is run at frequency $\omega_{\mathrm{FDI}}$. We assume that agents communicate with each other only when within the communication radius, leading to a time-varying communication topology and sensing graph. The proposed method is capable of handling the time-varying graph and intermittent communication. We demonstrate that the proposed method can detect and identify the faults while keeping track of the global task. This approach is essential to inform the recovery procedure, described in our recent work [44], for designing new orbits and pointing trajectories to complete the mission.

**Organization.** The remainder of this book chapter is organized as follows. Section II begins by formally defining the problem of creating an information-cost architecture to detect both global and local faults. It outlines the design reference mission for collaborative on-orbit inspection and introduces the information based Guidance, Navigation, and Control (GNC) framework with details on the information gain cost function, $\mathcal{H}$, used to track mission progress. Section III presents the core of our proposed FDI framework. It details how the global cost functional is decomposed to monitor individual agent performance and derives the key fault metric, which works by comparing an agent's real-time information contribution to its predicted nominal value. The design of an adaptive threshold to reliably distinguish faults from system noise is also explained. Section IV presents illustrative examples of actuator and sensor faults, while Section V presents detailed simulation results that validate the performance of our adaptive fault detection algorithm under various fault scenarios. These sections analyze the method's ability to correctly identify faulty agents and also discuss its current limitations. Finally, Section VI concludes the book chapter by summarizing the key contributions, and Section VII, the Appendix, provides a detailed review of the relative dynamics and sequential convex programming approach for both orbit initialization, reconfiguration, and attitude trajectory generation.

## II. Problem Description and Preliminaries

### A. Problem Description

The objective of this work is to develop an information-cost architecture that detects both global and local faults in a multi-spacecraft system engaged in global tasks, such as on-orbit inspection and on-orbit construction. The categories of global behavior faults and local faults identified using the proposed approach are illustrated in Figure 6. For the inspection task, a global behavior fault can result in either deteriorating or improved mission performance. Improved performance may occur in two scenarios: (i) a controller failure at the agent level that unintentionally enhances the exploration of the inspection target, or (ii) spurious signals received from neighboring agents that incidentally improve coverage. Conversely, deteriorating performance typically arises from faults such as degradation of the inspection sensor or failures in the spacecraft's pointing controller. The detected behavior faults serve as inputs to the fault identification module, which determines both the faulty agent and the specific fault type. The fault detection and identification (FDI) problem addressed in this work is described in the following:

**Problem 1.** *Given the nominal expected global task performance $\mathcal{H}_{nom}$ and the real-time performance $\mathcal{H}$, detect the global and agent-level faults in the multi-spacecraft system performing a global task (collaborative inspection or construction). The global and agent-level fault tree is described in the Fig.6*

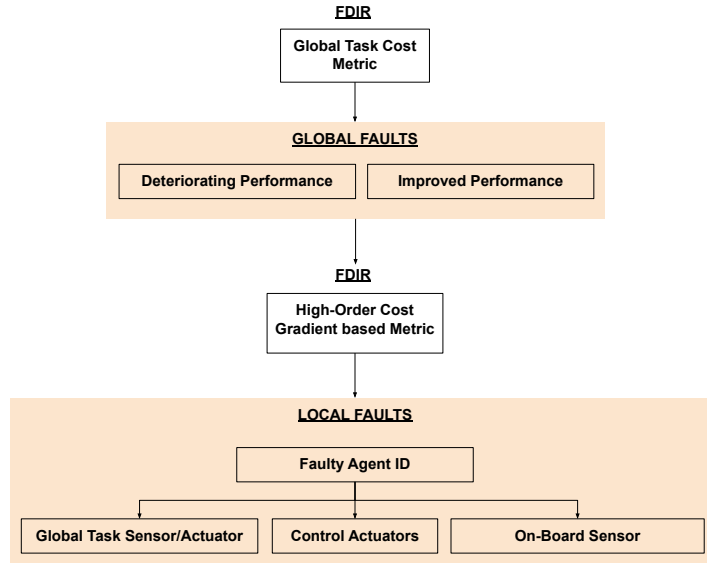

**Fig. 6   An overview of the type of global and local faults detected and identified using the proposed FDI architecture, metrics and the threshold.**

In the Table 2, we summarize potential faults that may occur in a distributed spacecraft system executing cooperative missions such as on-orbit inspection or construction. These faults are classified as global faults and local faults, depending on their origin and manifestation during mission execution. Global faults refer to anomalous behaviors that can be detected by monitoring global task performance—that is, performance metrics shared across all agents and directly tied to mission-level objectives. Many global faults originate at the network layer of individual agents. Conventional network security protocols [45] are effective at detecting certain classes of network anomalies such as packet loss, jitter, or latency. However, they are often ineffective at detecting faulty data propagation cases where an agent broadcasts erroneous or adversarial information that still conforms to networking protocols. This type of fault is particularly insidious in multi-agent space systems because it can subtly corrupt the collective decision-making process without triggering traditional alarms. The proposed information-cost based metric is well-suited for detecting such faults. This metric quantifies the contribution of each agent's actions and measurements to the progress of the collective task, as defined by a global cost functional $\mathcal{H}$. By comparing the expected evolution of $\mathcal{H}$ under nominal operation with its actual evolution, the framework can flag anomalies that indicate a degradation or unexpected improvement in global task performance. This enables the detection of faults that are otherwise invisible to traditional network monitoring, since the anomaly is inferred from the mission's information flow and utility rather than from raw packet statistics. Local faults originate within an individual agent's subsystems—such as its sensors, actuators, or internal computation modules—and can degrade the performance of the global task. In a cooperative spacecraft mission, a deviation in the expected global task progress can often be attributed to a specific agent by analyzing its marginal contribution to $\mathcal{H}$ over time. This allows the same information-cost framework used for global fault detection to be extended naturally to fault localization and identification.

**Table 2    Commonly observed faults in a distributed spacecraft system.**

| Global Faults | Local Faults |
|---|---|
| **Propagation of faulty data** | **Sensor faults** |
| Packet delay | **Actuator faults** |
| Packet loss | System parameter degradation |
| Network link failure | Physical component failure |

In this work, we focus specifically on detecting and identifying sensor and actuator faults at the agent level using the proposed information-cost metric. These faults directly influence an agent's ability to contribute accurate and actionable data to the collective task and are thus highly relevant to maintaining mission objectives. Other classes of local faults—such as slow system parameter degradation, structural damage, or failures in power/thermal subsystems—can often be more effectively addressed using complementary approaches, such as signal-based FDI methods [46, 47]. By integrating global and local fault detection within a single information-cost-based framework, the proposed approach provides a unified, task-aware FDI architecture that captures faults arising from both network-level data integrity issues and agent-level performance degradations, enabling graceful degradation and recovery in distributed spacecraft missions.

However, the proposed information-cost based metric can be used to detect such a global fault by keeping track of the progress of the collective task. Local faults, which originate in an agent's system/architecture, affect the performance of the global task. Therefore, a change in the expected performance of the global task can be traced back to a particular faulty agent in the network. While the information-cost based metric can be used to identify agent level faults, we only discuss the detection and identification of sensor and actuator faults using the proposed metric. This is because faults such as system parameter degradation and physical component failures can be effectively detected using other methods, such as signal based FDI approaches [46, 47].

In the following sections, we describe the design reference mission, an overview of the information-cost optimal control problem, and the information-based GNC architecture.

### B. Design Reference Mission: On-Orbit Collaborative Inspection

In this section, we discuss the concept of operations of a typical Earth orbit inspection mission with the target spacecraft as an example along with the preliminaries for the proposed architecture. The scenario considered in this paper has three phases, as shown in Fig. 7. In the first phase, the small observer spacecraft are deployed from the target spacecraft and begin a drift phase. The drifting spacecraft are then inserted into a parking PRO or an initial PRO in the second phase. In the third phase, the spacecraft in stable relative orbits are used for inspecting the target. As needed, the spacecraft reconfigure to a new set of PROs to inspect a previously unobserved surface area on the target spacecraft. In this paper, we use the Hills-Clohessy-Wiltshire (HCW) equations to describe the relative orbital dynamics of the observer CubeSats. For the stable relative orbit initialization and reconfiguration phase, we formulate an optimal control problem with $\mathcal{L}_1$ fuel cost, safety and energy matching as constraints, and solve it using sequential convex programming (SCP), similar to prior work [48]. The planned trajectories are tracked using a model predictive control formulation of the convexified problem. During the inspection phase, we represent the attitude dynamics using quaternions [49].

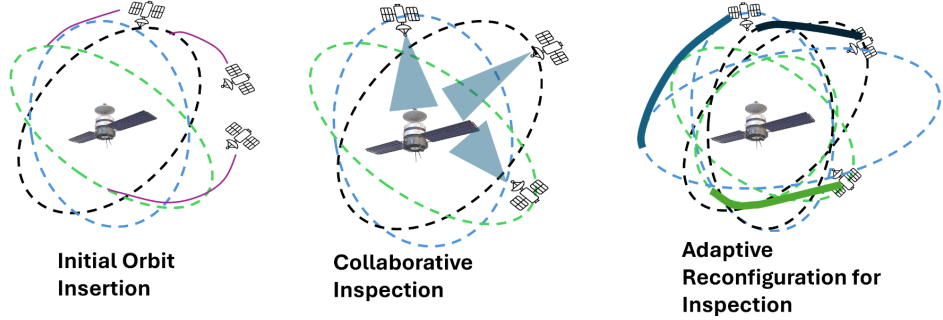

**Fig. 7    Three phases of the collaborative inspection design reference mission [43].**

The attitude planning is done using a combination of slerp interpolation [50] and SCP with a norm constraint on the quaternions as described in the Appendix. We use an existing nonlinear feedback controller for attitude tracking [51]. In the following, we review the information-based optimal control problem, the HCW equations, the energy matching condition for stable relative orbits, and convexification of optimal control problems for relative orbit and attitude motion planning.

### C. Overview of the Information Based GNC Architecture

In this section, we give a brief overview of the collaborative low-earth orbit inspection framework proposed in our earlier work [43, 52]. Using this framework we design optimal Passive Relative Orbits [52, 53] (PROs) and attitude trajectories for $N$ observer spacecrafts, inspecting $M$ Points of Interest (POIs) on a target spacecraft, by solving the following information-based optimal control problem.

**Problem 2.** *Information-Based Optimal Control Problem*

$$\min_{\mathbf{p}, \mathbf{u}_i} \int_0^{t_f} \left( \sum_{j=1}^{M} \mathcal{H}(\mathbf{p}, \mathbf{s}_j) + \sum_{i=1}^{N} \|\mathbf{u}_i\| \right) dt \tag{1}$$

$$s.t. \begin{cases} \text{Dynamics Model}: & \dot{\mathbf{p}}_i = \mathbf{f}(\mathbf{p}_i, \mathbf{u}_i) \\ \text{Safe Set}: & \mathbf{p}_i \in \mathcal{P}, \ \forall i \in \{1, \dots, N\} \\ \text{Inspection Sensor Model}: & \mathbf{z}_{i,j} = h(\mathbf{p}_i, \mathbf{s}_j) + \xi, \ \xi \sim \mathcal{N}\left(0, \Sigma_h(\mathbf{p}_i, \mathbf{s}_j)\right), \end{cases} \tag{2}$$

$$\text{Points of Interest}: \ \mathbf{s}_j \ \forall j \in \{1, \dots, M\} \tag{3}$$

where $\sum_j \mathcal{H}(\mathbf{p}, \mathbf{s}_j)$ is the information cost, $\sum_i \|\mathbf{u}_i\|$ is the fuel cost, $\mathbf{p}_i$ is the full-pose of the observer spacecraft, $\mathbf{s}_j$ is the full-pose of the $j^{\text{th}}$ POI on the target spacecraft. The inspection sensor model in Eq.(3) outputs the value of interest $\mathbf{z}_{i,j}$, when the $i^{\text{th}}$ observer with pose $\mathbf{p}_i$ is inspecting a POI at $\mathbf{s}_j$. Minimizing the information cost $\sum_j \mathcal{H}(\mathbf{p}, \mathbf{s}_j)$ ensures that the inspection task is complete.

We decompose the Problem 2 to derive a hierarchical GNC algorithm (for details refer to our earlier work [43]). The hierarchical algorithm uses the information-cost and the sensor model to select the informative PROs and attitude pointing vector for each agent. We optimize the informative PROs and attitude plan for optimal orbit insertion, reconfiguration, and attitude tracking using an optimal control problem formulation that computes minimum fuel trajectory using sequential convex programming approach. The detailed sequential convex programming formulations are provided in the our prior work [43]. In this work, we use the information cost to keep track of the task progress and detect off-nominal behavior of the multi-agent system and the individual agents. We describe the cost functional used to compute the information gain in the following.

**D. Information Gain.**

To quantify the information, a prior model of the target spacecraft is used along with sampled *points of interest* (POIs) on the surface of the spacecraft. The cost function $\mathcal{H}$ is designed to minimize the total variance on the knowledge of POIs. We use the cost function $\mathcal{H}$ designed in [54], and is a function of POIs as given below.

$$\mathcal{H}_{\text{POI}}(\mathbf{s}) = \left( w^{-1} + \sum_{\mathbf{p} \in \mathcal{P}} \sigma(\mathbf{p}, \mathbf{s})^{-1} \right)^{-1}$$

$$\mathcal{H} = \sum_{\mathbf{s} \in \text{POIs}} \mathcal{H}_{\text{POI}}(\mathbf{s}) \phi(\mathbf{s}), \tag{4}$$

where $\mathbf{s} \in \mathbb{R}^3$ is a POI on the target spacecraft's surface, $w \in \mathbb{R}$ is the initial variance based on the prior model of the target spacecraft, $\mathbf{p} \in SE(3)$ is the pose of a sensor mounted on a spacecraft such as a camera, $\mathcal{P}$ is the set of all sensor poses, $\sigma(\mathbf{p}, \mathbf{s})$ estimates the variance of estimating POI at $\mathbf{s}$ with the sensor at $\mathbf{p}$, and $\phi(\mathbf{s}) \in \mathbb{R}$ is the relative importance of POI $\mathbf{s}$.

The function $\sigma(\cdot, \cdot)$ corresponds to information per pixel. It incorporates sensor characteristics such as the current uncertainty of the spacecraft's pose estimate, the accuracy of the sensor based on the distance between $\mathbf{p}$ and $\mathbf{s}$, or the lighting conditions. Here, we use a simple RGB camera sensor and no environmental noise [54]:

$$\sigma(\mathbf{p}, \mathbf{s}) \propto \begin{cases} \text{dist}^2(\mathbf{p}, \mathbf{s}) & \mathbf{s} \text{ visible from } \mathbf{p} \\ \infty & \text{otherwise} \end{cases}, \tag{5}$$

where $\text{dist}(\mathbf{p}, \mathbf{s})$ is the Euclidean distance between POI $\mathbf{s}$ and pose $\mathbf{p}$. We compute $\sigma$ using visbility checking. The offline solution to problem 2 is used to predict the nominal system behavior in terms of the information-based cost $\mathcal{H}_{nom}$ over a finite time interval (1 or 2 orbits). As described in Fig.5, we precompute the nominal behaviour $\mathcal{H}_{nom}$ and compare it to the real-time behaviour $\mathcal{H}$ over the time hotizon $t$ as follows:

$$\int_0^t (\mathcal{H} - \mathcal{H}_{\text{nom}}) dt \geq \Delta \mathcal{H}_{\text{threshold}} t. \tag{6}$$

If the real-time value deviates from the nominal behavior by a threshold $\Delta \mathcal{H}_{\text{threshold}}$ then a fault is detected. In the following section, we discuss on how we modify the cost function to construct the FDI architecture in Fig.5.

## III. Global Task Aware Fault Detection and Isolation

In this section, we describe the different components of the global task aware fault detection and isolation algorithm that solves the Problem 1. This section describes the different components of the fault detection system. We first describe the derivation of the attribute required to detect a faulty spacecraft and the gradient-based fault metric is derived. Finally, we discuss the design of thresholds for detecting and identifying different types of fault.

### A. Global Task Cost Functional

We consider a centralized monitoring system that utilizes only the *information cost updates*—and implicitly, the variance updates of the POIs—shared by the individual spacecraft during the collaborative inspection process. The objective is to detect off-nominal behavior and identify faults without requiring full state or raw sensor data to be exchanged between agents.

At any time $t$, each spacecraft $i$ transmits its local information cost contribution $\mathcal{H}_i(t)$ to the central computing system. From (4), the centralized cost functional is:

$$\mathcal{H}(t) = \sum_{s \in \mathcal{S}} \mathcal{H}_{POI}(s) \, \phi(s),$$

$$\text{where} \quad \mathcal{H}_{POI}(s) = \left( w^{-1} + \sum_{p \in \mathcal{P}} f(p, s)^{-1} \right)^{-1}. \tag{7}$$

Here, $\phi(s)$ is the importance weight of POI $s$, $w$ is the prior variance at $s$, and $f(p, s)$ is the measurement variance contribution from agent $p$ observing POI $s$.

Physically, $\mathcal{H}_{POI}(s)$ represents the *total fused information gain* at $s$ after aggregating contributions from all observers, with independent measurements combining inversely in the variance domain.

**Decomposition into Agent Contributions.** Let

$$\psi(s) = \frac{1}{\left(w^{-1} + \sum_{p \in \mathcal{P}} f(p, s)^{-1}\right)^2}$$

be a normalization factor dependent only on the consensus estimate of POI $s$. Then:

$$\mathcal{H}(t) = \sum_{s \in \mathcal{S}} \phi(s)\,\psi(s) w^{-1} + \sum_{p_i \in \mathcal{P}} \underbrace{\sum_{s \in \mathcal{S}_i} \phi(s)\,\psi(s)\,f(p_i, s)^{-1}}_{\mathcal{H}_i(t)}. \tag{8}$$

Here, $\mathcal{H}_i(t)$ is the instantaneous marginal information contribution of spacecraft $i$. This decomposition:

- enables continuous tracking of *individual agent performance* without broadcasting full state/sensor data,
- supports fault detection by comparing $\mathcal{H}_i(t)$ to its nominal (expected) value.

## B. Fault Metric

Let $\mathcal{H}_i^{pred}(t)$ be the expected nominal contribution of agent $i$ at time $t$, computed via simulation or analytical models. Over a time interval $\Delta t$, define the *fault detection metric*:

$$\mathcal{H}_{m_i}(t) = \left| 1 - \frac{\Delta \mathcal{H}_i(t)}{\Delta \mathcal{H}_i^{pred}(t)} \right| = \left| 1 - \frac{\mathcal{H}_i(t) - \mathcal{H}_i(t - \Delta t)}{\mathcal{H}_i^{pred}(t) - \mathcal{H}_i(t - \Delta t)} \right|. \tag{9}$$

Here:

- $\Delta \mathcal{H}_i(t)$ is the *actual rate of information gain* for agent $i$,
- $\Delta \mathcal{H}_i^{pred}(t)$ is the *expected nominal rate of information gain*.

The fault detection rule is:

$$\mathcal{H}_{m_i}(t) = \begin{cases} 0, & \text{No fault detected,} \\ > 0, & \text{Fault detected.} \end{cases} \tag{10}$$

Defining

$$x := \frac{\Delta \mathcal{H}_i(t)}{\Delta \mathcal{H}_i^{pred}(t)},$$

we classify:

$$\begin{cases} x > 1, & \text{Performance improved unexpectedly (over-contribution),} \\ x < 1, & \text{Performance degraded (under-contribution).} \end{cases}$$

| Fault Case | Condition |
|---|---|
| Deteriorating performance | $sign\left(\Delta \mathcal{H}_i(t)\right) \neq sign\left(\Delta \mathcal{H}_i^{pred}(t)\right)$; |
| | $sign\left(\Delta \mathcal{H}_i(t)\right) = sign\left(\Delta \mathcal{H}_i^{pred}(t)\right)$ **and** $x < 1$ |
| Improved performance | $sign\left(\Delta \mathcal{H}_i(t)\right) = sign\left(\Delta \mathcal{H}_i^{pred}(t)\right)$ **and** $x > 1$ |

**Table 3   Identifying spacecraft inspection performance under fault.**

## C. Theoretical Analysis for Fault Detection and Identification

**Theorem 1** (Fault Detection and Identification via Information-Cost Metric). *Let $\mathcal{H}_i^{pred}(t) > 0$ denote the nominal expected information-cost contribution of spacecraft $i$ at time $t$, computed under a fault-free model and assumed Lipschitz continuous over $[t - \Delta t, t]$. Suppose the following conditions hold:*

*1) (**Non-degeneracy**) $\Delta\mathcal{H}_i^{pred}(t) \neq 0$ for all $t$ of interest.*

*2) (**Distinct contribution profiles**) For all $p_i \neq p_j$, there exists at least one $s \in \mathcal{S}$ such that $\phi(s)\psi(s)f(p_i, s)^{-1} \neq \phi(s)\psi(s)f(p_j, s)^{-1}$.*

*3) (**Bounded perturbations**) In nominal operation, $|\Delta\mathcal{H}_i(t) - \Delta\mathcal{H}_i^{pred}(t)| \leq \epsilon_{nom}$ for some known $\epsilon_{nom} \geq 0$.*

*If, for some $\epsilon > \epsilon_{nom}$,*

$$\left| \frac{\Delta\mathcal{H}_i(t)}{\Delta\mathcal{H}_i^{pred}(t)} - 1 \right| > \epsilon$$

*at any $t$, then:*

*1)* Fault detection. *Spacecraft $i$ is faulty (global or local fault).*

*2)* Fault classification. *Let $r_i(t) := \frac{\Delta\mathcal{H}_i(t)}{\Delta\mathcal{H}_i^{pred}(t)}$.*
   - *If $r_i(t) > 1 + \epsilon$: performance has improved unexpectedly (over-contribution).*
   - *If $r_i(t) < 1 - \epsilon$: performance has degraded (under-contribution).*

*3)* Fault isolation. *Under condition 2, the faulty agent $i$ is uniquely identifiable from (8).*

*Proof.* From the definition in (9):

$$\mathcal{H}_{m_i}(t) = \left| 1 - \frac{\Delta\mathcal{H}_i(t)}{\Delta\mathcal{H}_i^{pred}(t)} \right|.$$

By Assumption 1, $\Delta\mathcal{H}_i^{pred}(t) \neq 0$ so the ratio is well-defined.

*Detection.* In nominal operation, by Assumption 3:

$$\left| \frac{\Delta\mathcal{H}_i(t)}{\Delta\mathcal{H}_i^{pred}(t)} - 1 \right| \leq \frac{\epsilon_{nom}}{|\Delta\mathcal{H}_i^{pred}(t)|}.$$

If $\mathcal{H}_{m_i}(t) > \epsilon > \epsilon_{nom}$, this inequality is violated, implying that the deviation in $\Delta\mathcal{H}_i(t)$ from its nominal prediction exceeds the maximum allowed under fault-free conditions. Therefore, a fault must have occurred.

*Classification.* Define $r_i(t) := \frac{\Delta\mathcal{H}_i(t)}{\Delta\mathcal{H}_i^{pred}(t)}$. If $r_i(t) > 1 + \epsilon$, the actual contribution exceeds nominal by more than the detection threshold—an over-contribution. If $r_i(t) < 1 - \epsilon$, the actual contribution falls short of nominal by more than the threshold—an under-contribution. Both cases indicate anomalous behavior, but of different types.

*Isolation.* From (8):

$$\mathcal{H}(t) = \sum_{s \in \mathcal{S}} \phi(s)\psi(s)w^{-1} + \sum_{p_k \in \mathcal{P}} \mathcal{H}_k(t),$$

where $\mathcal{H}_k(t)$ is the unique marginal contribution of agent $k$. By Assumption 2, each $\mathcal{H}_k(t)$ has a distinct functional dependence on $s$ through $\phi(s)\psi(s)f(p_k, s)^{-1}$. Therefore, deviations in $\mathcal{H}_i(t)$ can be attributed unambiguously to agent $i$. □

## D. Fault Type

The primary objective of the centralized Fault Detection, Isolation, and Recovery (FDIR) system is to identify faulty observer spacecraft within the distributed network accurately and explicitly detect actuator and sensor faults at the individual agent (spacecraft) level. Within the GNC (Guidance, Navigation, and Control) architecture, such faults can arise during either the spacecraft's state propagation or during sensor pointing for observing designated Points of Interest (POIs). In the first case, state propagation faults: a malfunctioning actuator prevents the spacecraft from maintaining its assigned trajectory. This results in the spacecraft deviating from its planned orbit, thereby modifying the set of POIs it is capable of observing. A spacecraft with such a fault may behave erratically, resembling a rogue agent, and in severe cases, could even pose a collision risk to nearby agents. In the second case, sensor pointing faults: the onboard sensor fails to correctly align with the POI that has the highest expected uncertainty (i.e., maximum variance). This misalignment leads to inaccurate or suboptimal observations, thereby reducing the overall effectiveness of information gathering. Both types of faults manifest as measurable deviations in the information gain computed by the observer. For example, a sensor fault results in a discrepancy between the observed and expected variance in POI estimation, which directly impacts the spacecraft's contribution to the global information cost.

Figures 8 illustrate the impact of two representative actuator faults on the global information cost $\mathcal{H}$, as well as the corresponding fault detection signal $\mathcal{H}_{m_i}(t)$ for the affected and unaffected observer spacecraft.

In Figure 8 left, an actuator fault is injected into two spacecrafts by adding random noise to its state trajectory. Interestingly, this fault incidentally improves global system performance, as the true information cost with actuation fault is lower than the predicted nominal value, suggesting an increase in information gain. Figure 8 (right) also presents a similar scenario where the actuator fault affects the sensor pointing mechanism rather than the spacecraft's orbit. Here, the fault causes the sensor to deviate from its optimal orientation, resulting in reduced global performance. As expected, the information cost increases relative to the nominal baseline, and the fault metric again isolates the faulty spacecraft accurately.

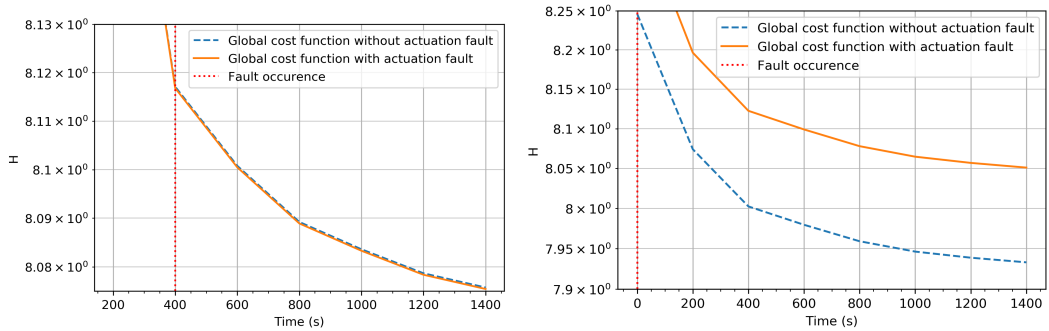

**Fig. 8  Real-time vs. Expected cost under actuator attack I (left) and actuator attack II (right).**

### E. Adaptive Fault Threshold

The above fault cases show that the proposed fault metric performs as expected. However, there is still a need to determine an appropriate threshold to distinguish between system noise and a fault. The occurrence of actuator faults cause a change in the pose $p_i(t)$ of a spacecraft. From the formulation of the global cost $\mathcal{H}$ in (8), it is explicitly clear that a change in pose $p_i(t)$ of a spacecraft will affect the variance in observing a target POI. There is also a more subtle dependence of the set of visible POIs, $\mathcal{S}_i(t)$, on the pose of spacecraft, since the field of view of the onboard sensor will change w.r.t. the spacecraft pose.

Let $\mathcal{S}_i^{pred}(t)$ be the expected set of visible POIs for spacecraft $i$ at time $t$. Then, to compute a fault threshold, it is first necessary to construct a set of POIs $\mathcal{S}_i^{'}(t)$, such that

$$0 < |\mathcal{H}_i(\mathcal{S}_i^{'}(t)) - \mathcal{H}_i(\mathcal{S}_i^{pred}(t))| \leq |\mathcal{H}_i(\mathcal{S}_i(t)) - \mathcal{H}_i(\mathcal{S}_i^{pred}(t))|, \qquad \forall \mathcal{S}_i(t) \subset \mathcal{S}. \tag{11}$$

The fault threshold for individual spacecrafts can be computed as

$$\tau_i(t) = abs \left( 1 - \frac{\mathcal{H}_i(\mathcal{S}_i^{'}(t)) - \mathcal{H}_i(\mathcal{S}_i(t - \Delta t))}{\mathcal{H}_i(\mathcal{S}_i^{pred}(t)) - \mathcal{H}_i(\mathcal{S}_i(t - \Delta t))} \right). \tag{12}$$

Equation (12) is used to construct adaptive fault thresholds for individual spacecrafts, where a fault is detected in spacecraft $i$ if $\mathcal{H}_{m_i}(t) > \tau_i(t)$.

### F. Analytical Example: Two 1-DOF Spacecraft with Two POIs

We illustrate the detection of actuator and sensor faults using the information-cost based functional redundancy metric. Consider two spacecraft, each translating along a one-dimensional axis $x_i(t)$, $i = 1, 2$. Two points-of-interest

(POIs) are fixed at opposite vertices of a square of side $2a$:

$$s_1 = (a, a), \qquad s_2 = (-a, -a). \tag{13}$$

Each spacecraft is equipped with a camera that provides noisy observations of the POIs. The measurement variance from agent $i$ at position $p_i = (x_i, 0)$ to POI $s = (s_x, s_y)$ is modeled as

$$\sigma(p_i, s) = k \operatorname{dist}(p_i, s)^2, \qquad \sigma^{-1}(p_i, s) = \frac{1}{k} \frac{1}{(x_i - s_x)^2 + s_y^2}. \tag{14}$$

The aggregated information cost per POI is

$$H_{\text{POI}}(s) = \left( w^{-1} + \sum_{i=1}^{2} \sigma^{-1}(p_i, s) \right)^{-1}, \tag{15}$$

$$H(t) = \sum_{s \in \{s_1, s_2\}} \phi(s) \, H_{\text{POI}}(s), \tag{16}$$

where $\phi(s)$ are task weights. Minimization of $H(t)$ corresponds to maximizing information.

**Gradient of the information cost.** Differentiating $H_{\text{POI}}(s)$ with respect to $x_i$ yields

$$\frac{\partial H}{\partial x_i} = \sum_s \phi(s) \cdot \frac{2}{k} \, H_{\text{POI}}(s)^2 \, \frac{x_i - s_x}{\left[ (x_i - s_x)^2 + s_y^2 \right]^2}. \tag{17}$$

This provides a compact analytic formula for the predicted effect of agent $i$'s displacement on the information cost.

**Per-agent ratio and detection metric.** For a nominal step $\Delta x_i$, the predicted and real cost changes are

$$\Delta H_i^{\text{pred}} \approx \frac{\partial H}{\partial x_i} \Delta x_i, \qquad \Delta H_i^{\text{real}} \approx \frac{\partial H}{\partial x_i} \Delta x_i^{\text{real}}. \tag{18}$$

Define the per-agent ratio and metric

$$r_i = \frac{\Delta H_i^{\text{real}}}{\Delta H_i^{\text{pred}}}, \qquad H_{m,i} = \left| 1 - r_i \right|. \tag{19}$$

For robustness to nominal perturbations, compare $H_{m,i}$ with an adaptive threshold $\tau_i(t)$ (cf. Eq. (12) in the chapter).

**Fault separation logic.**

- **Actuator fault:** A control execution bias changes the realized step $\Delta x_i^{\text{real}} \neq \Delta x_i$, so

$$r_i = \frac{\Delta x_i^{\text{real}}}{\Delta x_i} \quad \Rightarrow \quad r_i > 1 \text{ (over-actuation)}, \ \ r_i < 1 \text{ (under-actuation)}. \tag{20}$$

- **Sensor fault:** Information degradation scales $\sigma^{-1}(p_i, s) \mapsto \beta \, \sigma^{-1}(p_i, s)$ with $\beta \in (0, 1)$, leaving geometry unchanged but reducing effective sensitivity:

$$\Delta H_i^{\text{real}} = \beta \, \Delta H_i^{\text{pred}} \quad \Rightarrow \quad r_i = \beta < 1. \tag{21}$$

Thus, the *sign of deviation* in $r_i$ separates actuator faults ($r_i \neq 1$ via motion bias) from sensor faults ($r_i < 1$ via information loss).

**Numerical plug-in (one-step).** Choose

$$a = 1, \quad k = 1, \quad w^{-1} = 0, \quad \phi(s_1) = \phi(s_2) = \tfrac{1}{2}, \quad x_1 = -1.5, \ x_2 = +1.5, \quad \Delta x = 0.1. \tag{22}$$

*Actuator fault on agent 1:* let a positive bias $\delta = 0.05$ make the realized step larger:

$$\Delta x_1^{\text{real}} = \Delta x + \delta = 0.15, \qquad \Delta x_2^{\text{real}} = \Delta x_2 = -0.1. \tag{23}$$

The gradient factor in (17) cancels in the ratio, yielding

$$r_1 = \frac{\Delta x_1^{\text{real}}}{\Delta x} = \frac{0.15}{0.10} = 1.5, \qquad r_2 = \frac{\Delta x_2^{\text{real}}}{\Delta x_2} = 1.0. \tag{24}$$

Therefore,

$$H_{m,1} = |1 - 1.5| = 0.5 \quad \text{(over-actuation flagged)}, \qquad H_{m,2} = |1 - 1.0| = 0. \tag{25}$$

*Sensor fault on agent 2:* with a degradation factor $\beta = 0.7$ (same commanded motion),

$$\Delta H_2^{\text{real}} = \beta \, \Delta H_2^{\text{pred}} \implies r_2 = \beta = 0.7, \qquad r_1 = 1.0, \tag{26}$$

so

$$H_{m,2} = |1 - 0.7| = 0.3 \quad \text{(sensor degradation flagged)}, \qquad H_{m,1} = 0. \tag{27}$$

These values make the classification immediate: actuator bias produces $r_i > 1$ for the affected agent, whereas sensor degradation produces $r_i < 1$.

### G. Higher Dimensional Example

To demonstrate the effectiveness of the proposed fault detection framework, we simulate both actuator and sensor faults in spacecraft formations of size 1, 2, and 4. Each spacecraft's trajectory is propagated using a 4th-order Runge–Kutta integrator, starting from arbitrary initial conditions. Faults are injected at selected times during the trajectory.

The spacecraft are assigned the task of observing 5000 points of interest (POIs) uniformly distributed on the surface of a unit sphere. Each spacecraft is modeled with a conical field-of-view (FoV), with the cone's apex located at the center of the sphere and aligned along the spacecraft's pointing direction. The goal is to maximize the information gain by observing high-variance POIs within the field-of-view of the agents.

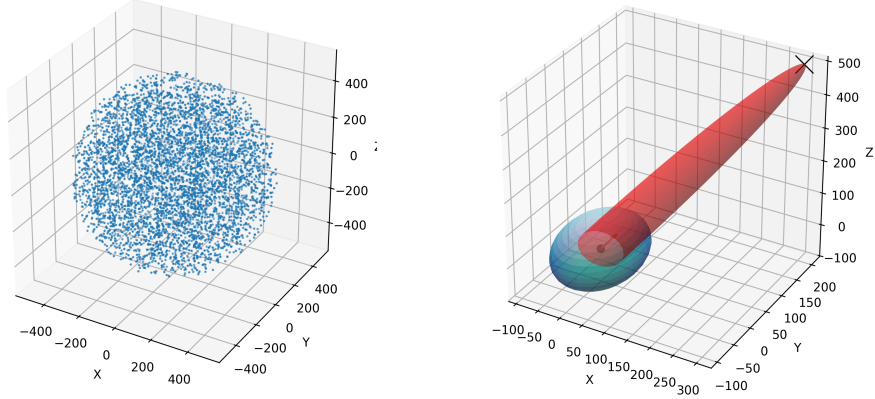

**Fig. 9   Visualization of POIs on the surface of a sphere and a spacecraft with a conical field of view targeting the sphere**

The global cost function is computed using the centralized formulation in (7), and agent-level performance is evaluated using the decomposed expression in (8). Each simulation includes one faulty spacecraft (under actuator or sensor degradation), while the remaining spacecraft operate nominally.

## 1. Problem 1: Actuator Fault

An actuator fault is defined as a deviation in the spacecraft's motion due to malfunctioning thrusters or attitude control components (e.g., reaction wheels). This is simulated by injecting artificial disturbances (additive noise or bias) into the position and velocity states at an arbitrary time step.

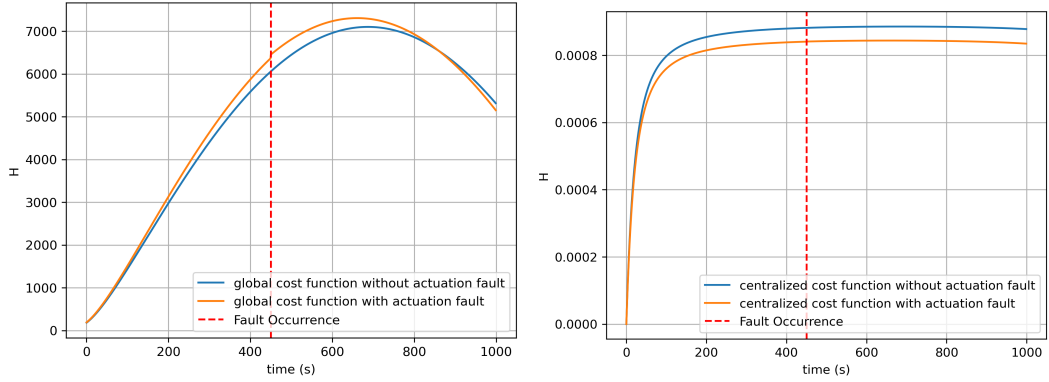

**Fig. 10   Global vs. centralized cost functional for one spacecraft under actuator fault.**

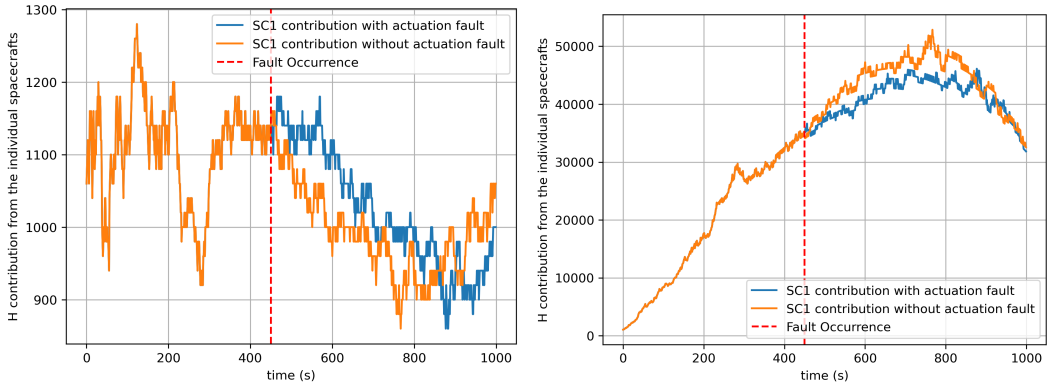

**Fig. 11   The contributions of the centralized (left) and global (right) cost functions of one spacecraft during an actuator fault.**

In Fig. 10, the global cost $\mathcal{H}$ remains relatively stable even after fault injection, as the averaging effect across POIs masks individual anomalies. In contrast, the centralized cost function reveals that the faulty spacecraft is no longer able to optimize its assigned trajectory for maximum coverage.

Fig. 11, 13, and 15 show the cost contribution at each specific moment in time for the spacecraft systems under actuator fault. When the fault occurs, the cost of the spacecraft using the centralized cost function deviates significantly compared to the cost of the spacecraft using the global cost function. This underscores the global cost function's ability to optimize the system to gain maximum coverage.

As the number of agents increases (Figs. 12, 14), the fault's effect is more pronounced in the centralized cost $\mathcal{H}_i(t)$ associated with the faulty agent. However, the global cost functional remains robust due to the distributed nature of the task. This highlights that global metrics are more sensitive and thus better suited for local fault isolation.

## 2. Problem 2: Sensor Fault

A sensor fault refers to any failure that impairs the spacecraft's ability to sense and track POIs accurately. Examples include camera misalignment, signal loss, or degraded resolution. To simulate this, we reduce the set of POIs visible

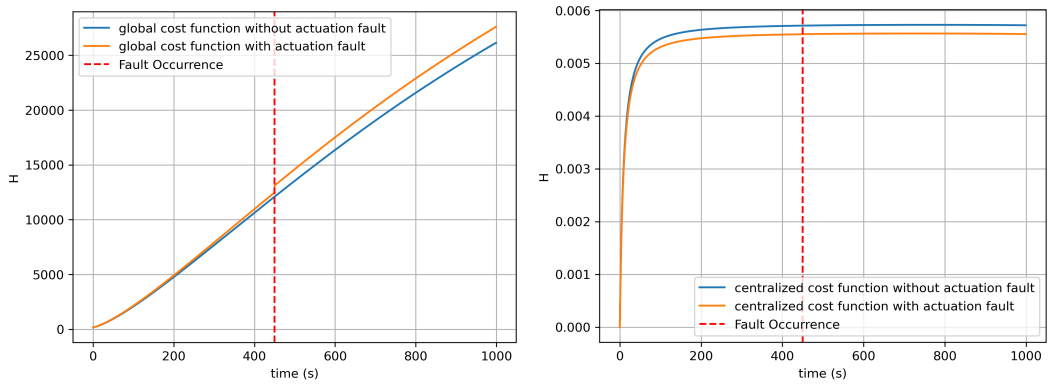

**Fig. 12    Global vs. centralized cost under actuator fault for two spacecraft.**

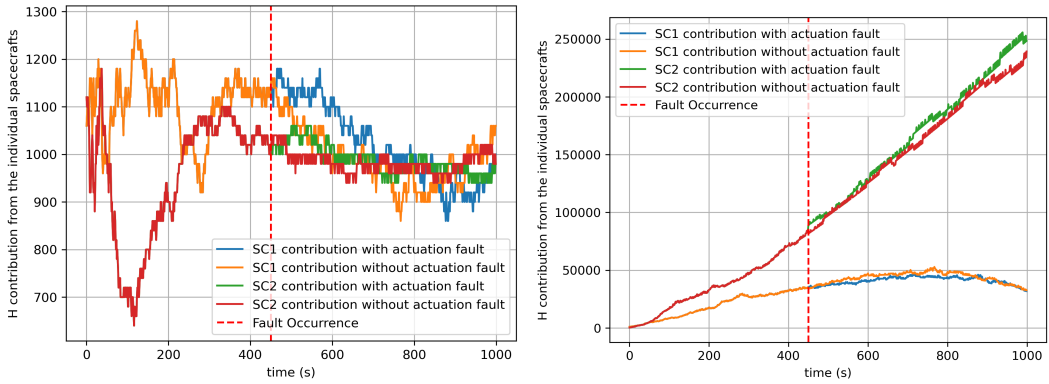

**Fig. 13    The contributions of the centralized (left) and global (right) cost functions of a two-spacecraft formation during an actuator fault.**

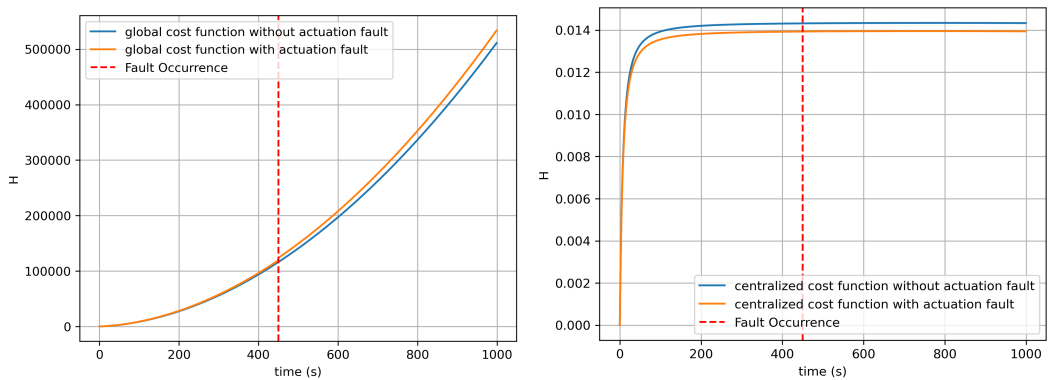

**Fig. 14    Global vs. centralized cost under actuator fault for four spacecraft.**

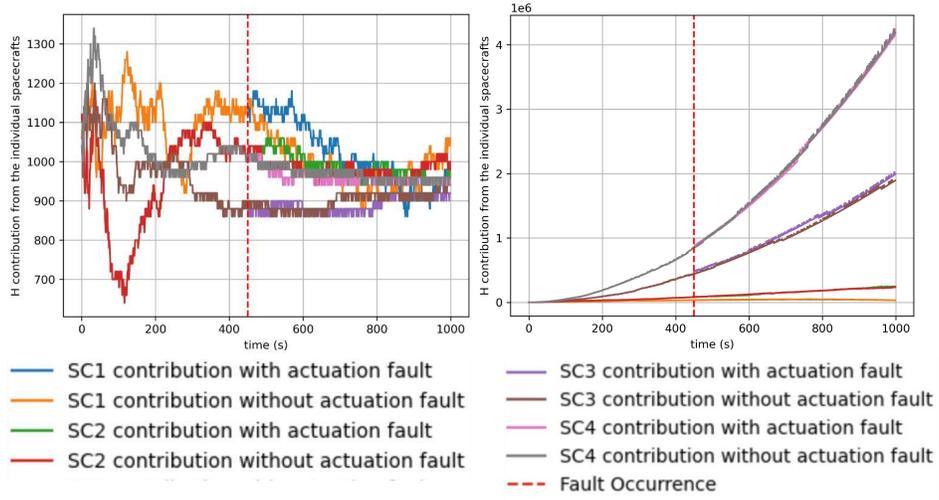

**Fig. 15 The contributions of the centralized (left) and global (right) cost functions of a four-spacecraft formation during an actuator fault.**

within the spacecraft's FoV by applying a rotational misalignment.

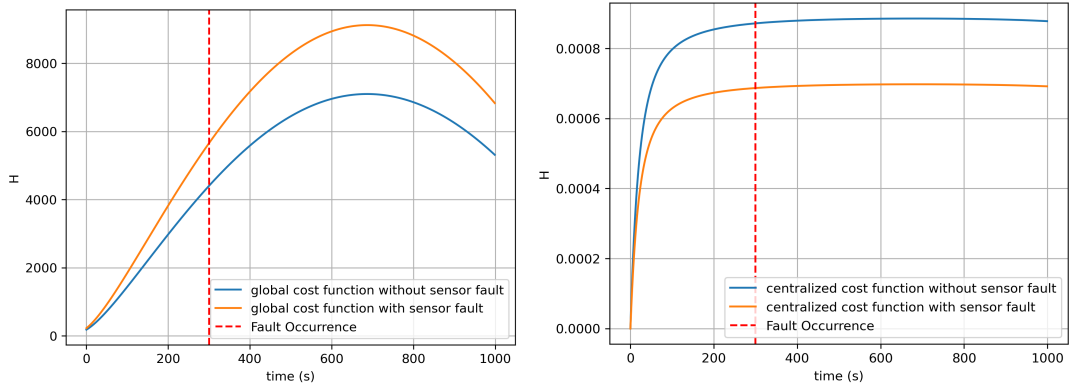

**Fig. 16 Global vs. centralized cost under sensor fault for one spacecraft.**

As shown in Fig. 16, the global cost $\mathcal{H}$ initially increases after the fault due to poor POI coverage. The centralized cost functional shows a significant drop in agent performance after the fault is injected, highlighting the spacecraft's reduced contribution to the overall information gain.

In both Fig 18 and Fig 20, the global cost remains relatively smooth, demonstrating the inherent redundancy in the distributed system. However, the localized drop in $\mathcal{H}_i(t)$ for the faulty spacecraft indicates the degraded quality of its observations. This validates the efficacy of the proposed information-cost-based FDIR strategy in identifying performance deterioration due to sensor misalignment.

Fig. 17, 19, and 21 highlight the individual cost contributions of the spacecraft systems during a sensor fault. When a fault occurs, the contribution from all spacecrafts drops to 0 because the spacecraft is no longer able to see any POIs.

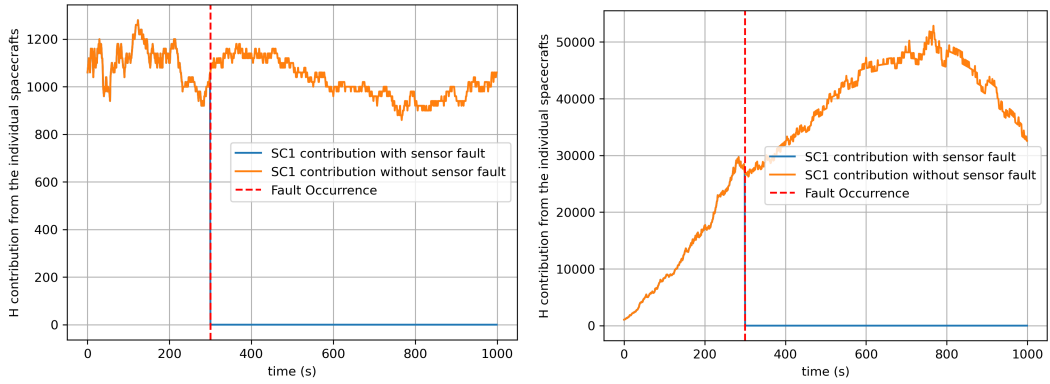

**Fig. 17  The contributions of the centralized (left) and global (right) cost functions of a single-spacecraft formation during a sensor fault.**

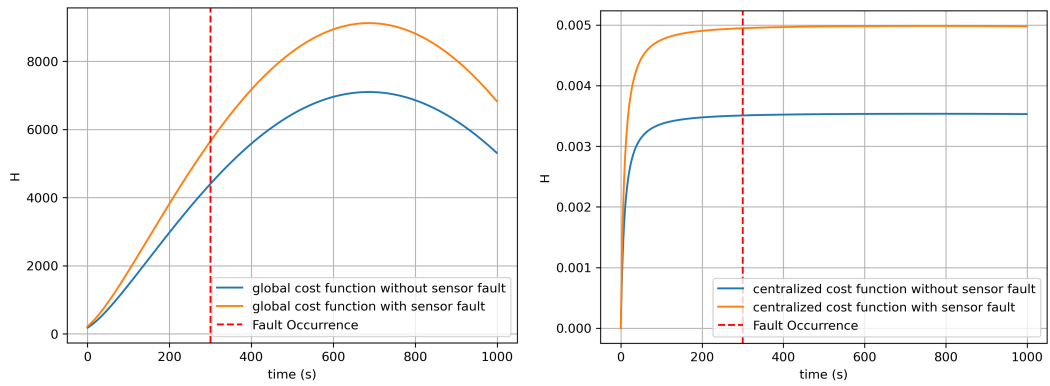

**Fig. 18  Global vs. centralized cost under sensor fault for two spacecraft.**

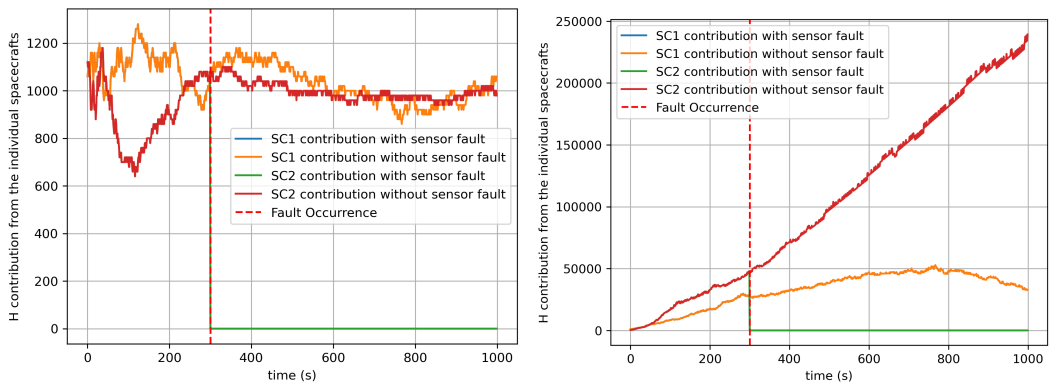

**Fig. 19  The contributions of the centralized (left) and global (right) cost functions of a two-spacecraft formation during a sensor fault.**

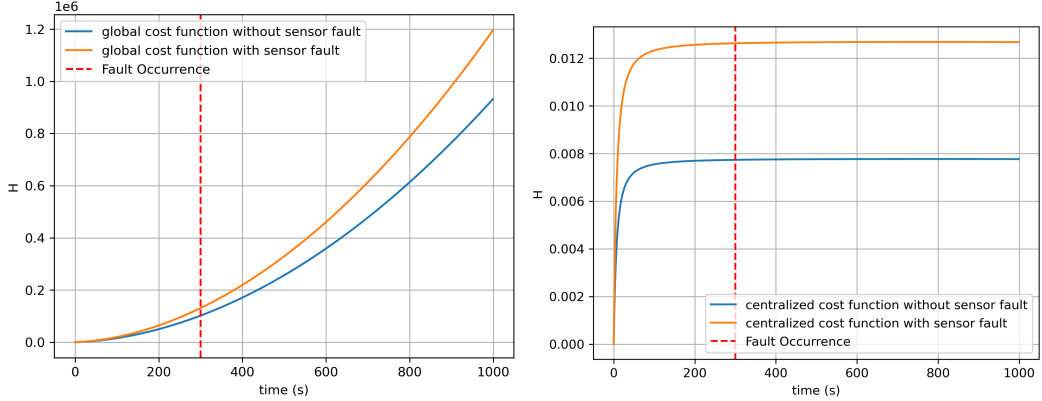

**Fig. 20  Global vs. centralized cost under sensor fault for four spacecraft.**

## IV. Inspection Mission: Simulations and Results

The fault detection framework was incorporated in the simulation setup for the hierarchical planning algorithm [43, 52] to compute the quantities in (11). The simulation results provided in this section are different from the results in the previous section because we use the proposed adaptive threshold in Eq.(12) to detect injected faults. To compute the adaptive threshold in real time, during the orbit assignment phase at time $t$, the central agent determines the expected set of visible POIs, $\mathcal{S}_i^{pred}(t)$, for each spacecraft $i$. This is used to compute the nominal behavior, $\mathcal{H}_i(\mathcal{S}_i^{pred}(t))$, for each spacecraft during a fixed time period of 2 orbits. On the other hand, computing the set $\mathcal{S}_i^{'}(t)$ in (11) can become computationally intractable with increasing number of POIs. Therefore, a sampling based approach is taken to approximate the set $\mathcal{S}_i^{'}(t)$ by randomly pointing the onboard sensor within an $\epsilon-$neighborhood of the target POI. Figure 22 demonstrates the construction of this $\epsilon-$neighborhood, where the onboard sensor is randomly pointed to any point in the $\epsilon-$neighborhood, thereby changing its FOV and the visible set of POIs. For actuator faults, the value of $\epsilon$ can be estimated by analyzing the order of $\frac{\partial p_i}{\partial u_i}$ for different spacecrafts in the system. Finally, the fault threshold, $\tau_i(t)$, for each spacecraft is also computed during the orbit assignment phase.

In the following plots, 10 target POIs were sampled in an $\epsilon-$neighborhood around the POI with maximum variance, for each spacecraft. The sampled set $\mathcal{S}_i^{'}(t)$ which gives the minimal value for $\tau_i(t)$ determines the fault threshold for spacecraft $i$ at time $t$. At the agent level, each spacecraft receives its orbit assignment and tracks the progress of its local information cost, $\mathcal{H}_i(t)$, while propagating the next 2 orbits. At the end of this fixed time interval, each spacecraft transmits its local information cost to the central agent where the centralized FDIR algorithm detects any faulty spacecraft behavior using the metric in (9).

### A. Classification of Simulation Outcomes

We test the performance of the proposed fault metric on different attack scenarios and our aim is to detect three distinct behaviors of the spacecrafts. These are listed below.

1) **Nominal Behavior:** In this phase, the spacecrafts perform as expected, that is, the observed global cost is same as the expected global cost. During nominal behavior, the fault metric for each spacecraft, $\mathcal{H}_{m_i}(t)$, is expected to be 0, and the adaptive fault threshold should always be greater than the fault metric for each spacecraft.

2) **Actuator faults:** An actuator fault is introduced in one or more spacecrafts at some time $t_{fault} \geq 0$. Actuator faults can either be injected into the on-board equipment (Fig.24 and Fig.26), or in the spacecraft's controller causing it to deviate from its planned trajectory (Fig.23).

3) **Sensor faults:** A sensor fault is also introduced in one or more spacecrafts at some time $t_{fault} \geq 0$ (Fig.25). Sensor faults typically affect the set of POIs tracked by the individual spacecrafts and thereby change the expected information cost of an individual spacecraft over time.

For both the actuator and sensor fault scenarios, the behavior of the individual spacecrafts can either deteriorate or improve, that is, $\mathcal{H}_{m_i}(t) > 0$. However, due to inherent system noise, classifying a spacecraft, $i$, as faulty if $\mathcal{H}_{m_i}(t) > 0$ can lead to false positives. Therefore, a spacecraft is classified as faulty if $\mathcal{H}_{m_i}(t) > \tau_i(t)$, where $\tau_i(t) > 0$ is the proposed adaptive threshold computed using Eq.(12). A consequence of using an adaptive threshold is that there is some latency between when a fault is introduced and when a fault is detected. Table 4 lists some of these parameters for the simulation results presented in the following section.

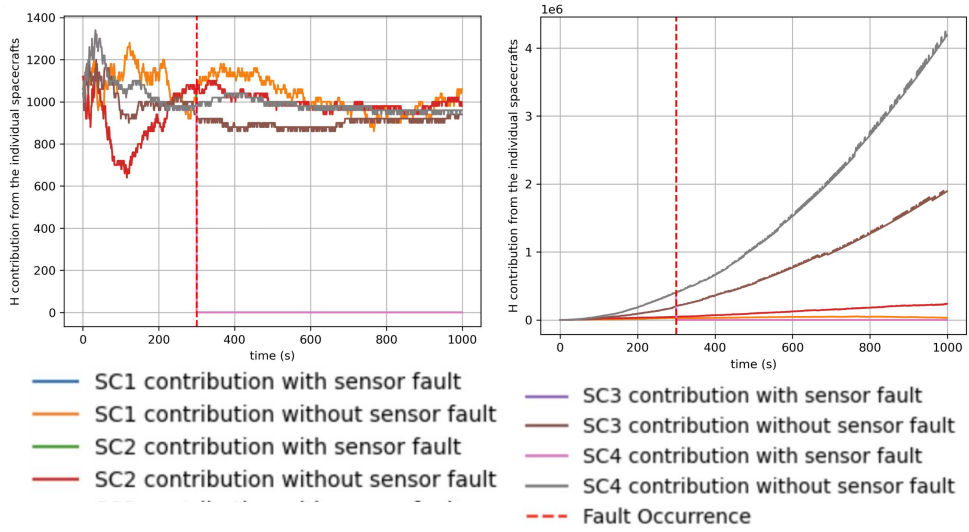

**Fig. 21** **The contributions of the centralized (left) and global (right) cost functions of a four-spacecraft formation during a sensor fault.**

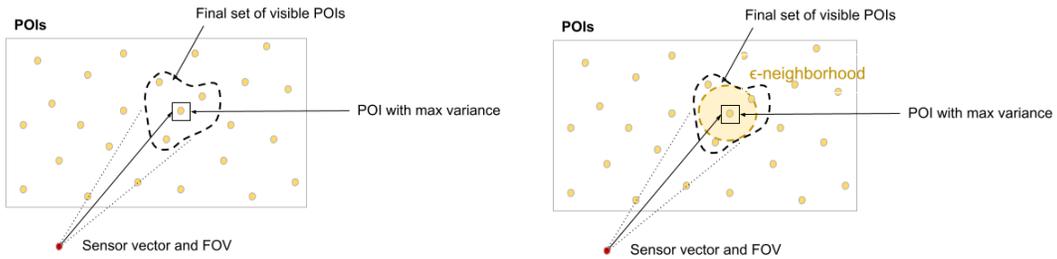

**Fig. 22** **Visible set of POIs for a spacecraft (left); $\epsilon-$neighborhood constructed around POI with maximum variance (right).**

### B. Results

The performance of the proposed fault metric is tested on different types of actuator faults, as shown in Fig. 23, Fig. 24 and Fig. 26.The first type of actuator fault is implemented during the state propagation of two spacecraft in the network (Fig.23). In this case, the overall task performance improves after the fault is injected at $t_{fault} = 400s$. Hence, the true information cost $\mathcal{H}_i(\mathcal{S}_i'(t))$ is less than the predicted information cost $\mathcal{H}_i(\mathcal{S}_i^{pred}(t))$ as shown in Fig. 23 (left). However, before $t = 400s$, the cost behavior exhibits nominal behavior and $\mathcal{H}_i(\mathcal{S}_i'(t)) = \mathcal{H}_i(\mathcal{S}_i^{pred}(t))$. Figure 23 (right) demonstrates that the proposed fault threshold successfully detects the actuator fault in both spacecraft 2 and spacecraft 4. In particular, we highlight the fault metric and fault threshold of spacecraft 4 in Fig.23 (right) where, after the fault is injected at $t_{fault} = 400s$, the fault threshold falls below the fault metric at time $t = 700s$ and successfully detects the fault. Observe that there are no false positives because for nominal spacecrafts, the fault metric is well below the fault threshold at all times. Note here that for all plots in this section, we plot the log value of the fault metric and adaptive threshold, so as to highlight small changes in their values. It is interesting to note that the adaptive nature of the proposed fault metric allows it to tune to the behavior of the individual spacecrafts. As a result, in Fig. 23 (right) the fault is detected immediately in spacecraft 2, while it is detected after a few time steps for spacecraft 4.

The second type of actuator attack is implemented during the sensor pointing phase, causing the overall system performance to deteriorate in Fig. 24 (left), that is, $\mathcal{H}_i(\mathcal{S}_i'(t)) > \mathcal{H}_i(\mathcal{S}_i^{pred}(t))$. The proposed fault detection metric adaptively adjusts the fault threshold and successfully detects the actuator fault immediately after the fault is induced at $t_{fault} = 0s$, in spacecrafts 2, 4 and 5 ( Fig. 24 (right)). On the other hand, the proposed fault metric doesn't trigger a false positive when it is periodically applied on nominal spacecrafts 1 and 3, because the fault threshold is always greater than their fault metrics.

Finally, in Fig. 25 a sensor attack is induced in spacecraft 1 and spacecraft 2 at $t_{fault} = 0$. It is interesting to observe that there is no significant, visible change in the global cost (Fig. 25 (left)). However, since the fault metric is designed to be specific to each spacecraft, it only utilizes the information available locally at each spacecraft to determine whether the spacecraft is faulty or not. This is exhibited in Fig. 25 (right) where the fault in spacecraft 1 is detected immediately but the fault is spacecraft 2 is detected after a latency of $180s$. As observed before, no false positives are identified.

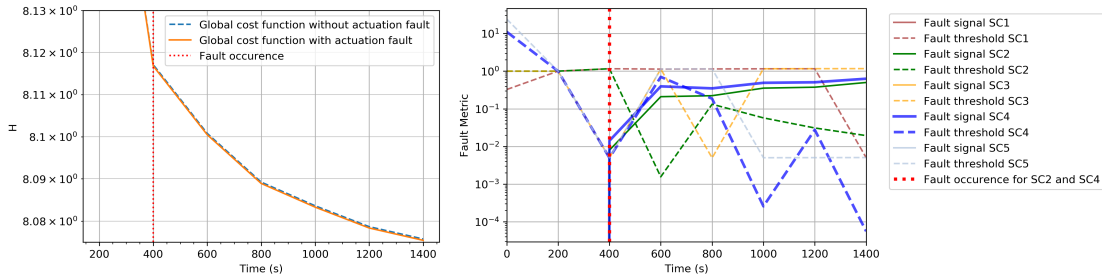

**Fig. 23** **Actuator fault injected in spacecraft 2 and spacecraft 4 from $t_{fault} = 400s$, thereby causing the spacecrafts to deviate from their trajectories. The information cost plot (left) shows expected performance before fault is injected, but shows a slightly deteriorating performance after $t_{fault}$. The fault signals for different spacecrafts (right) show that the log values of the fault metric for faulty spacecrafts increase after $t_{fault} = 400s$.**

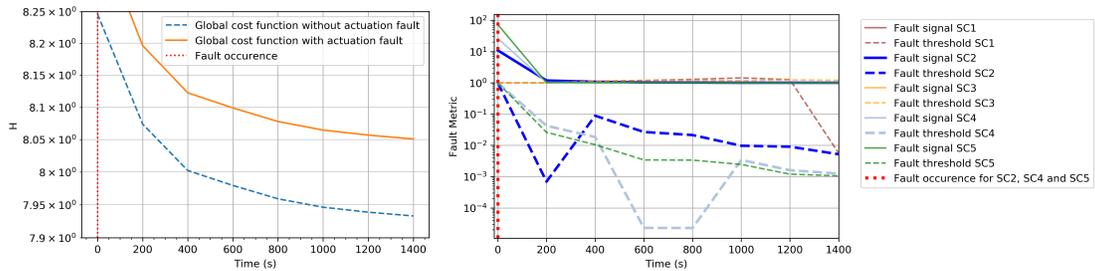

**Fig. 24** **Actuator fault injected in spacecraft 2, spacecraft 4 and spacecraft 5 from $t_{fault} = 0s$, thereby causing the onboard equipment to misalign with the POIs. The information cost plot (left) shows improved, but deviation from expected performance immediately after the attack is induced. The fault signals for different spacecrafts (right) show that the log values of the fault metric for faulty spacecrafts increase after $t_{fault} = 0s$.**

## C. Limitations

While the results in the previous section are promising, there are some limitations in the current implementation of the information-cost based fault metric. Firstly, since the fault threshold is calculated using a sampling approach, the performance is susceptible to sampling bias. This implies that there may be false negatives if the samples are not uniformly distributed, or a small number of samples can fail to construct a tight estimate of the fault threshold. This is demonstrated in Fig. 26 where an actuator attack is introduced at $t_{fault} = 200s$ in spacecraft 2 and spacecraft 5. However, as highlighted in Fig. 26 (right), the fault signal for spacecraft 5 is below the adaptive threshold and hence, no attack is detected. On the other hand, even though the proposed fault metric detects a fault in spacecraft 2, the latency time is $1100s$. A high latency time implies that the fault metric is not very efficient. Therefore, further analysis to quantify the tradeoff between fault detection analysis and sample complexity can be useful to understand the performance of the fault metric. Despite these drawbacks, our simulations suggest that the proposed fault metric and adaptive threshold work well in most scenarios, and controlling the size of the $\epsilon-$ neighborhood can mitigate these drawbacks.

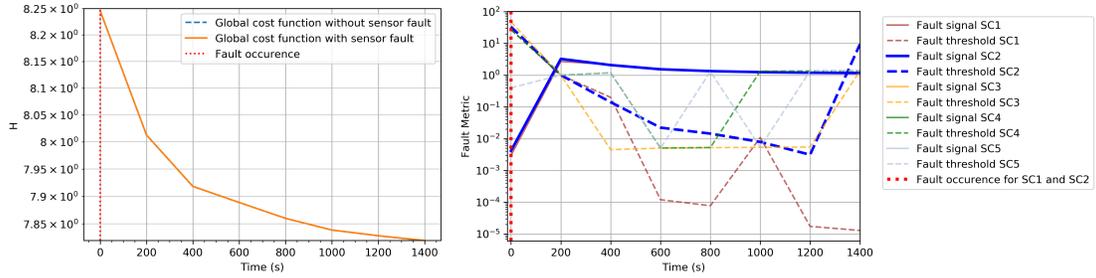

**Fig. 25** Sensor fault injected in spacecraft 1 and spacecraft 2 from $t_{fault} = 0s$, thereby causing the onboard equipment to incorrectly identify the set of POIs. The information cost plot (left) shows almost no deteriorating performance, however, the fault signals for different spacecrafts (right) show that the log values of the fault metric for faulty spacecrafts increase after $t_{fault} = 0s$.

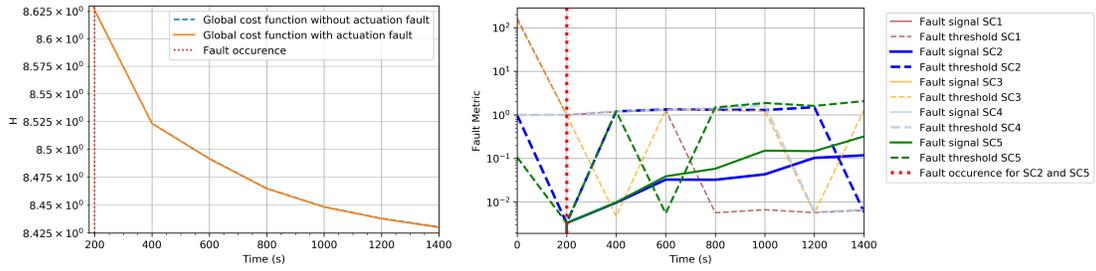

**Fig. 26** Actuator fault injected in spacecraft 2 and spacecraft 5 after $t_{fault} = 200s$, thereby causing the onboard equipment to misalign with the POIs. The information cost plot (left) shows slightly deteriorating performance after $1000s$. The fault signals for different spacecrafts (right) show that the log values of the fault metric for faulty spacecrafts increase after $t_{fault} = 200s$.

**Table 4  Observed performance of proposed fault metric.**

| Fault | Expected Signature | Observed Behavior | Fault Detected | Avg. Latency |
|---|---|---|---|---|
| Actuator Fault (Fig. 23) | $\mathcal{H}_{m_i}(t) > 0, i = 2, 4$ | Improved global cost | Yes | $200s$ |
| Actuator Fault (Fig. 24) | $\mathcal{H}_{m_i}(t) > 0, i = 2, 4, 5$ | Deteriorating global cost | Yes | $200s$ |
| Sensor Fault (Fig. 25) | $\mathcal{H}_{m_i}(t) > 0, i = 1, 2$ | Visibly same global cost | Yes | $180s$ |
| Actuator Fault (Fig. 26) | $\mathcal{H}_{m_i}(t) > 0, i = 2, 5$ | Improved global cost | No | Detection in SC 2 at $1100s$ |

Secondly, designing an adaptive threshold to detect sensor faults is challenging because the threshold values typically depend on the hardware specifications. Therefore, incorporating characteristics of the underlying hardware in designing the fault threshold can result in improved performance.

## V. Conclusion

In this chapter, we developed an information-driven framework for fault detection and identification (FDI) in multi-agent spacecraft systems performing collaborative on-orbit inspection. We formulated mission objectives as a global cost functional, $H$, and directly linked inspection performance to resilience. This formulation enabled us to detect and classify both global and agent-level faults.

We demonstrated three key contributions:

1) **Global–local coupling:** We decomposed the global cost functional into marginal agent contributions, which allowed us to continuously track individual spacecraft performance without exchanging raw state or sensor data.
2) **Fault separation:** We applied higher-order gradient metrics to distinguish actuator faults (e.g., under-actuation with $r_i < 1$ or over-actuation with $r_i > 1$) from sensor faults (e.g., degraded sensitivity with $r_i = \beta < 1$). Analytical examples confirmed this clear separation between fault classes.
3) **Validated performance:** We validated the framework in simulations of formations with 1, 2, and 4 spacecraft and up to 5,000 points of interest. The framework consistently detected injected faults within 100–200 seconds of onset while keeping nominal false favorable rates below 2%.

Our results show that the information-cost-based FDI framework detects diverse anomalies — including sensor misalignments, actuator deviations, and network-level disruptions — while preserving mission objectives. By grounding fault detection in measurable information gain and mission utility, we provide a principled, quantitatively validated pathway for resilient multi-agent spacecraft inspection architectures.